\documentclass[seceq]{ptptex}
\usepackage{wrapft}
\usepackage{graphicx}
\def\be{\begin{equation}}
\def\ee{\end{equation}}
\def\bea{\begin{eqnarray}}
\def\eea{\end{eqnarray}}

\pubinfo{Vol.~111, No.~4, April 2004}%Editorial Office will fill
\title{%        %You can use \\ for explicit line-break
String Gas Cosmology
}
\subtitle{\LaTeXe\ Version}    %use this when you want a subtitle

\author{%       %Use \scshape  for the family name
Robert H. \textsc{Brandenberger}\footnote{E-mail: rhb@hep.physics.mcgill.ca} 
}

\inst{%         %Affiliation, neglected when [addenda] or [errata]
Physics Department, McGill University, Montreal, QC, H3T 2A8, Canada}

\recdate{%      %Editorial Office will fill in this.
August 4, 2008}

\abst{%         %this abstract is neglected when [addenda] or [errata]
String gas cosmology is a string theory-based approach to early universe cosmology
which is based on making use of robust features of string theory such
as the existence of new states and new symmetries. A first goal of string gas
cosmology is to understand how string theory can effect the earliest moments of
cosmology before the effective field theory approach which underlies standard and
inflationary cosmology becomes valid. String gas cosmology
may also provide an alternative to the current standard paradigm of cosmology,
the inflationary universe scenario. Here, the current status of string gas cosmology
is reviewed.
}

\begin{document}

\maketitle

\section{Introduction}
\label{ch1:sec1}

\subsection{The Current Paradigm of Early Universe Cosmology}

According to the inflationary universe scenario \cite{Guth} (see also
\cite{Sato,Starob1,Brout}), there was a phase
of accelerated expansion of space lasting at least 50 Hubble expansion times
during the very early universe. This accelerated expansion of space can explain
the overall homogeneity of the universe, it can explain its large size and entropy,
and it leads to a decrease in the curvature of space. Most importantly, however,
it includes a causal mechanism for generating the small 
amplitude fluctuations which can
be mapped out today via the induced temperature fluctuations of the cosmic microwave
background (CMB) and which develop into the observed large-scale structure 
of the universe \cite{Mukhanov} (see also \cite{Press,Sato,Starob2,Lukash}).
The accelerated expansion of space stretches fixed co-moving scales beyond
the Hubble radius. Thus, it is possible to have a causal mechanism which
generates the fluctuations on microscopic sub-Hubble scales. The wavelengths of 
these inhomogeneities are subsequently inflated to cosmological scales which are 
super-Hubble until the late universe. The generation mechanism is based 
on the assumption that the fluctuations start out on
microscopic scales at the beginning of the period of inflation in a quantum
vacuum state. If the expansion of space is almost 
exponential, an almost scale-invariant spectrum of cosmological perturbations 
results, and the squeezing which the fluctuations undergo while they evolve on 
scales larger than the Hubble radius predicts a characteristic oscillatory pattern in
the angular power spectrum of the CMB anisotropies \cite{Sunyaev}, a pattern
which has now been confirmed with great accuracy \cite{Boomerang,WMAP}
(see e.g \cite{MFB} for a comprehensive review of the theory of cosmological
fluctuations, and \cite{RHBrev2} for an introductory overview).

To establish our notation, we write the metric of a homogeneous, isotropic
and spatially flat four-dimensional universe in the form
\be \label{metric}
ds^2 \, = \, dt^2 - a(t)^2 d{\bf x}^2 \, ,
\ee
where $t$ is physical time, ${\bf x}$ denote the three co-moving spatial
coordinates (points at rest in an expanding space have constant co-moving
coordinates), and the scale factor $a(t)$ is proportional to the size of
space. The expansion rate $H(t)$ of the universe is given by
\be
H(t) \, = \, \frac{{\dot a}}{a} \, ,
\ee
where the overdot represents the derivative with respect to time.

\begin{figure}
\begin{center}
%\centerline{\epsfxsize=3in\epsfbox{canc1.eps}}
\includegraphics[height=9cm]{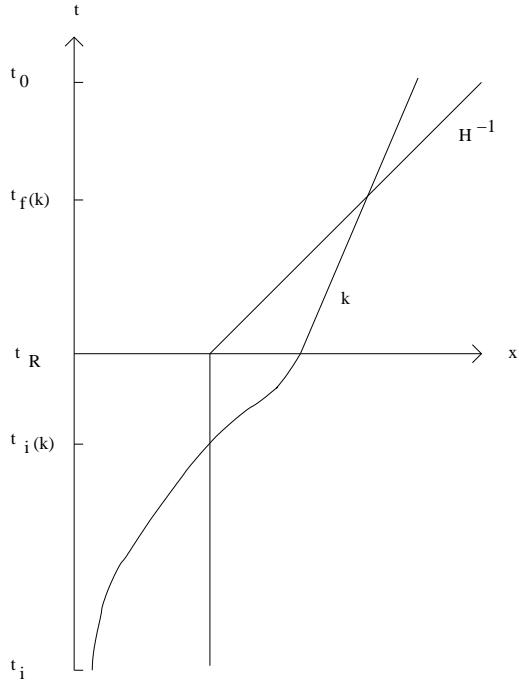}
\caption{Space-time diagram (sketch) 
of inflationary cosmology. Time increases along the vertical
axis. The period of inflation begins at time $t_i$, ends at
$t_R$, and is followed by the radiation-dominated phase
of standard big bang cosmology. If the expansion of space is exponential,
the Hubble radius $H^{-1}$ is constant in physical spatial coordinates
(the horizontal axis), whereas it increases linearly in time
after $t_R$. The physical length corresponding to a fixed
co-moving length scale is labelled by its wave number $k$ and increases
exponentially during inflation but increases less fast than
the Hubble radius (namely as $t^{1/2}$), after inflation. Hence, the 
wavelength crosses the Hubble radius twice. It exits the Hubble radius
during the inflationary phase at the time $t_i(k)$ and re-enters during
the period of standard cosmology at time $t_f(k)$.}
\end{center}
\label{fig:1}       
\end{figure}

A space-time sketch of inflationary cosmology is shown in Fig. 1. The vertical
axis is time. The inflationary phase begins at the time $t_i$ and lasts until
the time $t_R$, the time of ``reheating". At that time, the energy which is driving
inflation must change its form into regular matter. The Hubble radius is labelled
by $H^{-1}(t)$ and divides scales into those where micro-physics dominates and thus
the generation of fluctuations by local physics is possible (sub-Hubble scales)
and those where gravity dominates and micro-physical effects are negligible
(super-Hubble). As shown in the sketch, during inflation fixed co-moving
scales (labelled by $k$ in the sketch) are inflated from microscopic to
cosmological. Note also that the horizon, the
forward light cone, becomes exponentially larger than the Hubble radius
during the inflationary phase.

\subsection{Challenges for String Cosmology}

Working in the context of General Relativity as the theory of space-time, 
inflationary cosmology requires the presence of a new form of matter with
a sufficiently negative pressure $p$ ($p < - 2/3 \rho$, where $\rho$ denotes
the energy density). In order to obtain such an equation of state, in
general the presence of scalar field matter must be assumed. In addition, it
must be assumed that the scalar field potential energy dominates over
the scalar field spatial gradient and kinetic energies for a sufficiently long
time period. The Higgs field used for the spontaneous breaking of gauge
symmetries in particle physics has a potential which is not flat enough to
sustain inflation. Models beyond the Standard Model of particle physics, in
particular those based on supersymmetry, typically have many scalar fields.
Nevertheless, it has proven to be very difficult to construct viable inflationary 
models. The problems which arise when trying to embed inflation into
the context of effective field theories stemming from superstring theory
are detailed in the contribution to this book by Burgess.

If inflationary cosmology is realized in the context of classical General 
Relativity coupled to scalar field matter, then an initial cosmological
singularity is unavoidable \cite{Borde}. Resolving this initial singularity
is one of the challenges for string cosmology.

The energy scale during inflation is set by the observed amplitude of the
CMB fluctuations. In simple single field models of inflation, the energy
scale is of the order of the scale of Grand Unification, i.e. many orders
of magnitude larger than scales for which field theory has been tested
experimentally, and rather close to the string and Planck scales, scales
where we know that the low energy effective field theory approach will
break down. It is therefore a serious concern whether the inflationary
scenario is robust towards the inclusion of non-perturbative stringy
effects, effects which we know must not only be present but in fact must
dominate at energy scales close to the string scale. 

The problem for cosmological fluctuations is even more acute: provided
that the inflationary phase lasts for more than about 70 Hubble expansion
times, then all scales which are currently probed in cosmological
observations had a wavelength smaller than the Planck length at the
beginning of the inflationary phase. Thus, the modes definitely
are effected by trans-Planckian physics during the initial stages of their
evolution. The ``trans-Planckian problem" for fluctuations 
\cite{RHBrev1,Jerome} is whether
the stringy effects which dominate the evolution in the initial stages
leave a detectable imprint on the spectrum of fluctuations. To
answer this question one must keep in mind that the expansion of
space does not wash out specific stringy signatures, but simply
red-shifts wavelengths. For string theorists, the above ``trans-Planckian
problem" is in fact a window of opportunity: if the universe underwent a
period of inflation, this period will provide a microscope with which
string-scale physics can be probed in current cosmological observations.

Some of the conceptual problems of inflationary cosmology are
highlighted in Figure 2, a space-time sketch analogous to that
of Figure 1, but with the two zones of ignorance (length scales
smaller than the Planck (or string) length and densities higher
than the Planck (or string) density) are shown. As the
string scale decreases relative to the Planck scale, the
horizontal line which indicates the boundary of the
super-string density zone of ignorance approaches the 
constant time line corresponding to the onset of inflation.
This implies that the inflationary background dynamics itself might
not be robust against stringy corrections in the dynamical equations.

The sketch in Figure 2 also
shows the exponential increase of the horizon compared to the
Hubble radius during the period of inflation.

\begin{figure}
\begin{center}
%\centerline{\epsfxsize=3in\epsfbox{canc1.eps}}
\includegraphics[height=9cm]{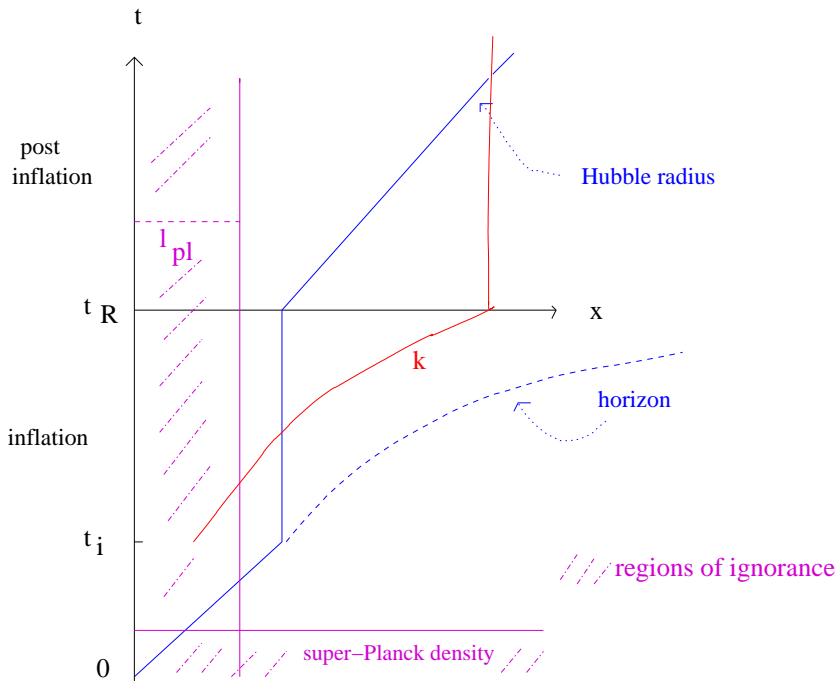}
\caption{Space-time diagram (sketch) 
of inflationary cosmology including the two zones of
ignorance - sub-Planckian wavelengths and trans-Planckian 
densities. The symbols have the same meaning as in Figure 1.
Note, specifically, that - as long as the period of inflation
lasts a couple of e-foldings longer than the minimal value
required for inflation to address the problems of standard
big bang cosmology - all wavelengths of cosmological interest
to us today start out at the beginning of the period of inflation
with a wavelength which is in the zone of ignorance.}
\end{center}
\label{fig:2}       
\end{figure}

\subsection{Preview}

The conceptual problems of inflationary cosmology discussed in the
previous subsection motivate a search for a new paradigm of early
universe cosmology based on string theory. Such a new paradigm may
provide the initial conditions for a robust inflationary phase. However, it
may also lead to an alternative scenario. In the following, we will
explore this second possibility. 

In the best possible world, the initial phase of string cosmology will
eliminate the cosmological ``Big Bang" singularity, it will provide a
unified description of space, time and matter, and it will allow a
controlled computation of the induced cosmological perturbations.
The development of such a consistent framework of string cosmology
will, however, have to be based on a consistent understanding of
non-perturbative string theory. Such an understanding is at the
present time not available. 

Given the lack of such an understanding,
most approaches to string cosmology are based on treating matter
using an effective field theory description motivated by string theory.
However, in such approaches key features of string theory which are
not present in field theory cannot be seen. The approach to string
cosmology discussed below is, in contrast, based on studying
effects of new degrees of freedom and new symmetries which
are key ingredients to string theory, which will be present in any
non-perturbative formulation of string theory.

\section{Basics of String Gas Cosmology}
\label{ch1:genrem}

\subsection{Principles of String Gas Cosmology}

In the absence of a non-perturbative formulation of string theory,
the approach to string cosmology which we have suggested,
{\it string gas cosmology} \cite{BV,TV,ABE} (see also \cite{Perlt},
and \cite{RHBrev6,BattWatrev} for reviews),
is to focus on symmetries and degrees of freedom which are new to
string theory (compared to point particle theories) and which will
be part of any non-perturbative string theory, and to use
them to develop a new cosmology. The symmetry we make use of is
{\bf T-duality}, and the new degrees of freedom are the {\bf string
oscillatory modes} and the {\bf string winding modes}.

String gas cosmology is based on coupling a classical background
which includes the graviton and the dilaton fields to a gas of
strings (and possibly other basic degrees of freedom of
string theory such as ``branes"). All dimensions of space are taken
to be compact, for reasons which will become clear later.
For simplicity, we take all spatial directions to be toroidal and
 denote the radius of the torus by $R$. Strings have three types
of states: {\it momentum modes} which represent the center
of mass motion of the string, {\it oscillatory modes} which
represent the fluctuations of the strings, and {\it winding
modes} counting the number of times a string wraps the torus.

Since the number of string oscillatory states increases exponentially
with energy, there is a limiting  temperature for a gas of strings in
thermal equilibrium, the {\it Hagedorn temperature} \cite{Hagedorn}
$T_H$. Thus, if we take a box of strings and adiabatically decrease the box
size, the temperature will never diverge. This is the first indication that
string theory has the potential to resolve the cosmological singularity
problem (see also \cite{KalRam,Easson03} for discussions on how the 
temperature singularity can be avoided in string cosmology).

The second key feature of string theory upon which string gas cosmology
is based is {\it T-duality}. To introduce this symmetry, let us discuss the
radius dependence of the energy of the basic string states:
The energy of an oscillatory mode is independent of $R$, momentum
mode energies are quantized in units of $1/R$, i.e.
\be
E_n \, = \, n {1 \over R} \, ,
\ee
and winding mode energies are quantized in units of $R$, i.e.
\be
E_m \, = \, m R \, ,
\ee
where both $n$ and $m$ are integers. Thus, a new symmetry of
the spectrum of string states emerges: Under the change
\be
R \, \rightarrow \, 1/R
\ee
in the radius of the torus (in units of the string length $l_s$)
the energy spectrum of string states is
invariant if winding
and momentum quantum numbers are interchanged
\be
(n, m) \, \rightarrow \, (m, n) \, .
\ee
The above symmetry is the simplest element of a larger
symmetry group, the T-duality symmetry group which in
general also mixes fluxes and geometry.
The string vertex operators are consistent with this symmetry, and
thus T-duality is a symmetry of perturbative string theory. Postulating
that T-duality extends to non-perturbative string theory leads
\cite{Pol} to the need of adding D-branes to the list of fundamental
objects in string theory. With this addition, T-duality is expected
to be a symmetry of non-perturbative string theory.
Specifically, T-duality will take a spectrum of stable Type IIA branes
and map it into a corresponding spectrum of stable Type IIB branes
with identical masses \cite{Boehm}. 

As discussed in \cite{BV}, the above T-duality symmetry leads to
an equivalence between small and large spaces, an equivalence
elaborated on further in \cite{Hotta,Osorio}.

\subsection{Dynamics of String Gas Cosmology}

That string gas cosmology will lead to a dynamical evolution of the
early universe very different from what is obtained in standard and
inflationary cosmology can already be seen by combining the
basic ingredients from string theory discussed in the previous
subsection. As the radius of a box of strings decreases from an
initially very large value - maintaining thermal
equilibrium - , the temperature first rises as in
standard cosmology since the string states which are occupied
(the momentum modes) get heavier. However, as the temperature
approaches the Hagedorn temperature, the energy begins to
flow into the oscillatory modes and the increase in temperature
levels off. As the radius $R$ decreases below the string scale,
the temperature begins to decrease as the energy begins to
flow into the winding modes whose energy decreases as $R$
decreases (see Figure 3). Thus, as argued in \cite{BV},  
the temperature singularity of early universe cosmology
should be resolved  in string gas cosmology.

\begin{figure}
\begin{center}
%\centerline{\epsfxsize=3in\epsfbox{canc1.eps}}
\includegraphics[height=6cm]{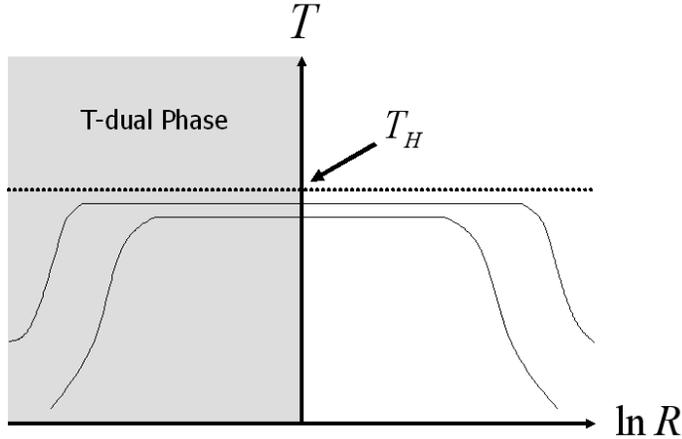}
\caption{The temperature (vertical axis) as a function of
radius (horizontal axis) of a gas of closed strings in thermal
equilibrium. Note the absence of a temperature singularity. The
range of values of $R$ for which the temperature is close to
the Hagedorn temperature $T_H$ depends on the total entropy
of the universe. The upper of the two curves corresponds to
a universe with larger entropy.}
\end{center}
\label{fig:3}       
\end{figure}

The equations that govern that background of string gas cosmology
are not known. The Einstein equations are not the correct
equations since they do not obey the T-duality symmetry of
string theory. Many early studies of string gas cosmology were
based on using the dilaton gravity equations \cite{TV,Ven,Tseytlin}. However,
these equations are not satisfactory, either. Firstly,
we expect that string theory correction terms to the
low energy effective action of string theory become dominant
in the Hagedorn phase. Secondly, the dilaton gravity
equations yields a
rapidly changing dilaton during the Hagedorn phase (in the
string frame). Once the dilaton becomes large, it becomes
inconsistent to focus on fundamental string states rather
than brane states. In other words, using dilaton gravity as a
background for string gas cosmology does not correctly
reflect the S-duality symmetry of string theory. Recently, a
background for string gas cosmology including a rolling
tachyon was proposed \cite{Kanno1} which allows a background
in the Hagedorn phase with constant scale factor and constant
dilaton. Another study of this problem was given in \cite{Sduality}.

Some conclusions about the time-temperature relation in string
gas cosmology can be derived based on thermodynamical
considerations alone. One possibility is that  $R$ starts out
much smaller than the self-dual value and increases monotonically.
From Figure 3 is then follows that the time-temperature curve
will correspond to that of a bouncing cosmology. Alternatively,
it is possible that the universe starts out in a meta-stable state
near the Hagedorn temperature, the {\it Hagedorn phase}, and
then smoothly evolves into an expanding phase dominated by
radiation like in standard cosmology (Figure 4). Note that we
are assuming that not only is the scale factor constant in time,
but also the dilaton.  

\begin{figure}
\begin{center}
%\centerline{\epsfxsize=3in\epsfbox{canc1.eps}}
\includegraphics[height=6cm]{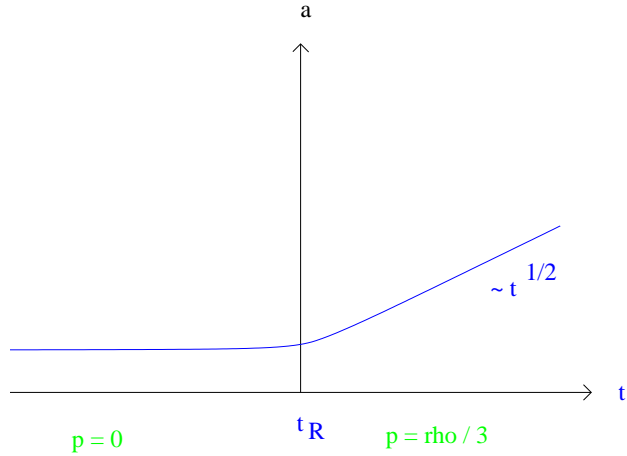}
\caption{The dynamics of string gas cosmology. The vertical axis
represents the scale factor of the universe, the horizontal axis is
time. Along the horizontal axis, the approximate equation of state
is also indicated. During the Hagedorn phase the pressure is negligible
due to the cancellation between the positive pressure of the momentum
modes and the negative pressure of the winding modes, after time $t_R$
the equation of state is that of a radiation-dominated universe.}
\end{center}
\label{fig:4}       
\end{figure}

The transition between the quasi-static Hagedorn phase and the
radiation phase at the time $t_R$ is a consequence of 
the annihilation of string winding modes into string loops (see Figure 5). 
Since this process corresponds to the production of radiation, we denote
this time by the same symbol as the time of reheating in inflationary
cosmology. As pointed out in \cite{BV}, this annihilation process
only is possible in at most three large spatial dimensions. This is
a simple dimension counting argument: string world sheets have
measure zero intersection probability in more than four large 
space-time dimensions. Hence, string gas cosmology
may provide a natural mechanism for explaining why there are
exactly three large spatial dimensions. This argument was
supported by numerical studies of string evolution in three and
four spatial dimensions \cite{Mairi} (see also \cite{Cleaver}). 
The flow of energy from
winding modes to string loops can be modelled by effective
Boltzmann equations \cite{BEK} analogous to those used to
describe the flow of energy between infinite cosmic strings and
cosmic string loops (see e.g. \cite{VilShell,HK,RHBrev4} for
reviews). 

Several comments are in place concerning the above mechanism.
First, in the analysis of \cite{BEK} it was assumed that the
string interaction rates were time-independent. If the dynamics of
the Hagedorn phase is modelled by dilaton gravity, the dilaton is
rapidly changing during the phase in which the string frame scale
factor is static. As discussed in \cite{Col2,Danos}, in this case the
mechanism which tells us that exactly three spatial dimensions 
become macroscopic does not work. Another comment concerns
the isotropy of the three large dimensions. In contrast to the
situation in Standard cosmology, in string gas cosmology the
anisotropy decreases in the expanding phase \cite{Watson1}.
Thus, there is a natural isotropization mechanism for the three
large spatial dimensions.

\begin{figure}
\begin{center}
%\centerline{\epsfxsize=3in\epsfbox{canc1.eps}}
\includegraphics[height=4.5cm]{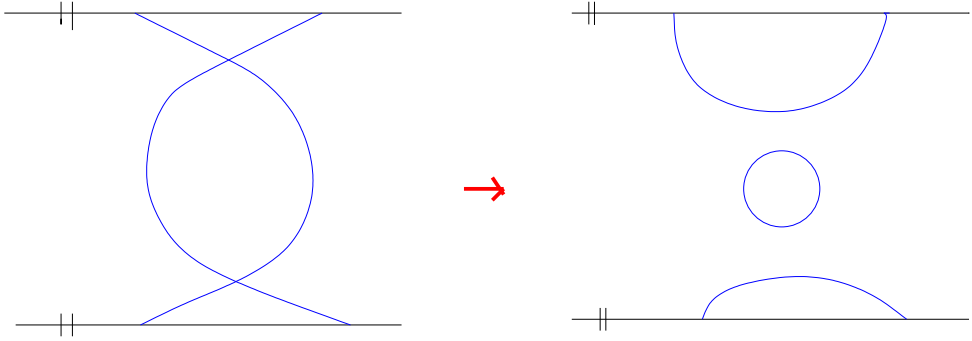}
\caption{The process by which string loops are produced via the
intersection of winding strings. The top and bottom lines
are identified and the space between these lines represents
space with one toroidal dimension un-wrapped.}
\end{center}
\label{fig:5}       
\end{figure}

At late times, the dynamics of string gas cosmology can be
described by dilaton gravity or - if the dilaton is fixed - by
Einstein gravity. The dilaton gravity action coupled to
string gas matter is
\be \label{dilgrav}
S \, = \, {1 \over {2 \kappa^2}} \int d^{10}x \sqrt{-g} e^{-2 \phi}
\bigl[{\hat R} + 4 \partial^{\mu} \phi \partial_{\mu} \phi \bigr] + S_m \, ,
\ee
where $g$ is the determinant of the metric, ${\hat R}$ is the Ricci scalar,
$\phi$ is the dilaton, 
$\kappa$ is the reduced gravitational constant in ten dimensions,
and $S_m$ denotes the matter action for which we will use the
hydrodynamical action of a string gas. The
metric appearing in the above action is the metric in the
string frame. 

In the case of a homogeneous
and isotropic background given by (\ref{metric})
the three resulting equations (the
generalization of the two Friedmann equations plus the equation
for the dilaton) in the string frame are
\cite{TV} (see also \cite{Ven})
\bea
-d {\dot \lambda}^2 + {\dot \varphi}^2 \, &=& \, e^{\varphi} E 
\label{E1} \\
{\ddot \lambda} - {\dot \varphi} {\dot \lambda} \, &=& \,
{1 \over 2} e^{\varphi} P \label{E2} \\
{\ddot \varphi} - d {\dot \lambda}^2 \, &=& \, {1 \over 2} e^{\varphi} E \, ,
\label{E3}
\eea
where $E$ and $P$ denote the total energy and pressure, respectively,
$d$ is the number of spatial dimensions, and we have introduced the
logarithm of the scale factor 
\be
\lambda(t) \, = \, {\rm log} (a(t))
\ee
and the rescaled dilaton
\be
\varphi \, = \, 2 \phi - d \lambda \, .
\ee
The above equations are consistent with a fixed dilaton in the radiation phase,
but not in the Hagedorn phase (see e.g. \cite{Betal}). As we run backwards in time,
the dilaton runs off towards a singularity which is inconsistent with the
ideas of a quasi-static Hagedorn phase. A detailed study of the dynamics
of the background space-time in the presence of string gases with both
Hagedorn and radiation equations of state was performed in \cite{Borunda}
\footnote{Corrections to these equations coming from stringy $\alpha^{\prime}$
terms were considered in \cite{Borunda2}.}.

This set of equations (\ref{E1}),(\ref{E2}),(\ref{E3}) can be supplemented 
with Boltzmann type equations
which describe the transfer of energy from the string winding modes
to string loops \cite{BEK}. The equations describe how
two winding strings with opposite orientations
intersect, producing closed loops with vanishing winding
as a final state (see Figure 5). First, we split the energy density in 
strings into the density in winding strings
\be
\rho_w(t) \, = \, \nu(t) \mu t^{-2} \, ,
\ee
where $\mu$ is the string mass per unit length, and $\nu(t)$ is
the number of strings per Hubble volume, and into the density in
string loops
\be
\rho_l(t) \, = \, g(t) e^{-3 (\lambda(t) - \lambda(t_0))} \, ,
\ee
where $g(t)$ denotes the co-moving number density of loops, normalized
at a reference time $t_0$. In terms of these variables, the
equations describing the loop production from the interaction of
two winding strings are \cite{BEK}
\bea 
{{d \nu} \over {dt}} \, &=& \, 
2 \nu \bigl( t^{-1} - H \bigr) - c' \nu^2 t^{-1} \label{E4} \\
{{dg} \over {dt}} \, &=& \, c' \mu t^{-3} \nu^2 
e^{3 \bigl( \lambda(t) - \lambda(t_0) \bigr)} \label{E5}
\eea
where $c'$ is a constant, which is of order unity for cosmic strings
but which depends on the dilaton in the case of fundamental strings
\cite{Col2,Danos}. 

If the spatial size is large in the Hagedorn phase, not all winding
strings will disappear at the time $t_R$. In fact, 
as is well known from the studies of cosmic strings 
\cite{VilShell,HK,RHBrev4},  the above
transfer equations (\ref{E4}),(\ref{E5}) lead to the existence of a scaling solution
for cosmic superstrings according to which at any given time in
the radiation phase for $t > t_R$, there will be a distribution of
cosmic superstrings characterized by a constant average
number of winding strings crossing each Hubble volume. A remnant distribution
of cosmic superstrings at all late times is thus one of the
testable predictions of string gas cosmology.

\section{Moduli Stabilization in String Gas Cosmology}

\subsection{Principles}

A major challenge in string cosmology is to stabilize all of the
string moduli. Specifically, the sizes and shapes of the extra
dimensions must be stabilized, and so must the dilaton. In string
gas cosmology based on heterotic superstring theory, all of the
size and shape moduli are fixed by the basic ingredients of the
model, namely the presence of string states with both momentum
and winding modes.  

The stabilization of the size moduli was
considered in \cite{Watson2,Subodh1,Subodh2,radion}, that of
the shape moduli in \cite{Edna,shape} (see \cite{RHBrev5} for
a review).  The basic principle is the following: in a string
gas containing both momentum and winding modes, the
winding modes will prevent expansion since their energies
increase with $R$ whereas the momentum modes will prevent
contraction since their energies scale as $1/R$. Thus, on
energetic grounds, there is a preferred value for the size of
the extra dimensions, namely $R = 1$ in string units. In
heterotic string theory, there are {\it enhanced symmetry states}
which contain both momentum and winding quantum numbers
and which are massless at the self-dual radius. These are
the lowest energy states near the self-dual radius and hence
dominate the thermodynamic partition function. These states
act as radiation from the point of view of our three large
dimensions, and are hence phenomenologically acceptable
at late times \cite{Subodh2}. The role of these states for
moduli stabilization was stressed in a more general context in 
\cite{Watson3,beauty,BrianScott}. 

It turns out that the shape moduli are also stabilized by the
presence of the enhanced symmetry states, without
requiring any additional inputs. The only modulus
which requires additional input for its stabilization is
the dilaton (the problem of simultaneously stabilizing
both the dilaton and the radion in the context of
dilaton gravity coupled to perturbative string theory
states was discussed in detail in \cite{Aaron}). 

\subsection{Stabilization of Geometrical Moduli}

The stabilization of the geometrical moduli at late times can be
analyzed in the context of dilaton gravity (the discussion
in this subsection is close to the one given in \cite{RHBrev5}). We use the
following ansatz for an anisotropic metric with scale factor 
$a(t) = exp(\lambda(t))$ for the three large dimensions and 
corresponding scale factor $b(t) = exp(\nu(t))$ for the internal 
dimensions (considered here to be isotropic):
\be
ds^2 \, = \, dt^2 - e^{2 \lambda} d{\bf x}^2 - e^{2 \nu} d{\bf y}^2 \, ,
\ee
where ${\bf x}$ are the coordinates of the three large dimensions and
${\bf y}$ the coordinates of the internal dimensions.

The variational equations of motion for $\lambda(t)$, $\nu(t)$ and $\phi(t)$
which follow from the dilaton
gravity action are \cite{Watson2}
\bea
- 3 {\ddot \lambda} - 3 {\dot \lambda}^2 - 6 {\ddot \nu} - 6 {\dot \nu}^2
+ 2 {\ddot \phi} \, &=& \, {1 \over 2} e^{2 \phi} \rho \\
{\ddot \lambda} + 3 {\dot \lambda}^2 + 6 {\dot \lambda} {\dot \nu}
- 2 {\dot \lambda} {\dot \phi} \, &=& \, {1 \over 2} e^{2 \phi} p_{\lambda} \\
{\ddot \nu} + 6 {\dot \nu}^2 + 3 {\dot \lambda} {\dot \nu}
- 2 {\dot \nu} {\dot \phi} \, &=& \, {1 \over 2} e^{2 \phi} p_{\nu} \\
- 4 {\ddot \phi} + 4{\dot \phi}^2 - 12 {\dot \lambda}{\dot \phi}
- 24 {\dot \nu} {\dot \lambda} + 3 {\ddot \lambda}  && \nonumber \\
+ 6 {\dot \lambda}^2 + 6 {\ddot \nu} + 21 {\dot \nu}^2 + 18 {\dot \lambda} {\dot \nu} \, 
&=& \, 0 \, .
\eea
where $\rho$ is the energy density and $p_{\lambda}$ and $p_{\nu}$
are the pressure densities in the large and the internal directions, 
respectively.

Let us now consider a superposition of several string gases, one with
momentum number $M_3$ about the three large dimensions, one with
momentum number $M_6$ about the six internal dimensions, and a further
one with winding number $N_6$ about the
internal dimensions. Note that there are no winding modes about
the large dimensions ($N_3 = 0$), either because they have already annihilated
by the mechanism discussed in the previous section, or they were never
present in the initial conditions. In this case, the energy $E$ and
the total pressures $P_{\lambda}$ and $P_{\nu}$ are given by
\bea
E \, &=& \, 
\mu \bigl[ 3  M_3 e^{- \lambda} + 6 M_6 e^{- \nu} + 6 N_6 e^{\nu} \bigr] \\
P_{\lambda} \, &=& \, \mu M_3 e^{- \lambda} \\ 
P_{\nu} \, &=& \, \mu \bigl[ - N_6 e^{\nu} + M_6 e^{- \nu} \bigr] \, ,
\label{EA3}
\eea
where $\mu$ is the string mass per unit length. Below,
we will consider a more realistic string gas, a gas
made up of string states which have momentum, winding and oscillatory
quantum numbers together. The states considered here are massive,
and would not be expected to dominate the thermodynamical partition
function if there are states which are massless. However, for the
purpose of studying radion stabilization in the string frame, the
use of the above naive string gas is sufficient.

We are interested in the symmetric case $M_6 = N_6$
In this case, it follows from (\ref{EA3}) that the equation of motion for
$\nu$ is a damped oscillator equation, with the minimum of the effective
potential corresponding to the self-dual radius. The damping is due
to the expansion of the three large dimensions (the expansion
of the three large dimensions is driven by the pressure
from the momentum modes $N_3$). Thus, we see that the
naive intuition that the competition of winding and momentum modes about
the compact directions stabilizes the radion degrees of freedom at the
self-dual radius generalizes to this anisotropic setting.

However, in the context of dilaton gravity, the dilaton is rapidly evolving in
the Hagedorn phase. Thus, the Einstein frame metric is not
static even if the string frame metric is (see e.g. \cite{BattWat}). The
key question is whether the radion remains stabilized if the dilaton
is fixed by hand (or by mechanisms discussed below). For a gas
of strings made up of massive states such as considered above this
is not the case. In Heterotic string theory, there are {\it enhanced
symmetry states} which are massless at the self-dual radius,
hence dominate the thermodynamic partition function, and can
stabilize the radion \cite{Subodh2}. In the following we will discuss
this mechanism. 

The equations of motion which arise from coupling string gas
matter to the Einstein (as opposed to the dilaton gravity) action
lead to - for an anisotropic metric of the form
\be
ds^2 \, = \, dt^2 - a(t)^2 d{\bf x}^2 - 
\sum_{\alpha = 1}^6 b_{\alpha}(t)^2 dy_{\alpha}^2 \, ,
\ee
where the $y_{\alpha}$ are the internal coordinates - the
following equation for the radion $b_{\alpha}$ 
\be \label{extra}
{\ddot b_{\alpha}} + 
\bigl( 3 H + \sum_{\beta = 1, \beta \neq \alpha}^6 {{{\dot b_{\beta}}} \over {b_{\beta}}} \bigr) 
{\dot b_{\alpha}} \, = \, 
\sum_{n, m} 8 \pi G {{\mu_{m,n}} \over {\sqrt{g} \epsilon_{m,n}}}{\cal S} \, .
\ee
The vector index pairs $(m, n)$ label perturbative string states. 
Note that $n$ and $m$ are momentum and winding
number six-vectors, one component for each internal dimension.
Also, $\mu_{m,n}$ is the number density of string states with the momentum
and winding number vector pair $(m, n)$, $\epsilon_{m,n}$
is the energy of an individual $(m,n)$ string, and $g$ is the determinant of
the metric. The source term ${\cal S}$ depends on the quantum numbers of the
string gas, and the sum runs over all  $m$ and $n$. If the number
of right-moving oscillator modes is given by $N$, then the source term
for fixed $m$ and $n$ is
\be \label{source}
{\cal S} \, = \, \sum_{\alpha} \bigl( {{m_{\alpha}} \over {b_{\alpha}}} \bigr)^2
- \sum_{\alpha} n_{\alpha}^2 b_{\alpha}^2 
+ {2 \over {D - 1}} \bigl[ (n,n) + (n, m) + 2(N - 1) \bigr] \, ,
\ee
where $(n, n)$ and $(n, m)$ indicate scalar products relative to the
metric of the internal space.
To obtain this equation, we have made use of the mass spectrum of
string states and of the level matching conditions. In the case of
the bosonic superstring, the mass spectrum for fixed $m, n, N$ and 
${\tilde N}$,
where ${\tilde N}$ is the number of left-moving oscillator states,
on a six-dimensional torus whose radii are given by $b_{\alpha}$ is
\be
m^2 \, = \, \sum_{\alpha} \bigl({{m_{\alpha}} \over {b_{\alpha}}} \bigr)^2
- \sum_{\alpha} n_{\alpha}^2 b_{\alpha}^2 + 2 (N + {\tilde N} - 2) \, ,
\ee
and the level matching condition reads
\be
{\tilde N} \, = \, (n,m) + N \, .
\ee

There are modes which are massless at the self-dual radius $b_{\alpha} = 1$.
One such mode is the graviton with $n = m = 0$ and $N = 1$. The modes of
interest to us are modes which contain winding and momentum, namely 
\begin{itemize}
\item{} $N = 1$, $(m,m) = 1$, $(m, n) = -1$ and $(n,n) = 1$;
\item{} $N = 0$, $(m,m) = 1$, $(m, n) = 1$ and $(n,n) =  1$;
\item{} $N = 0$  $(m,m) = 2$, $(m, n) = 0$ and $(n,n) =  2$.
\end{itemize}
The above discussion was in the context of bosonic string theory.
Due to the presence of the bosonic string theory tachyon, the above
states are not the lowest energy states for bosonic string theory
and hence do not dominate the thermodynamic partition function.
In Heterotic string theory, the tachyon is factored out of the
spectrum by the GSO \cite{Pol} projection, but the states
we discussed above survive. In contrast, in Type II string
theory, our massless states are also factored out. Thus, in
the following we will restrict attention to Heterotic string theory.

In string theories which admit massless states (i.e. states
which are massless at the self-dual radius), these states
will dominate the initial partition function. The background
dynamics will then also be dominated by these states. To understand
the effect of these strings, consider the equation of motion (\ref{extra})
with the source term (\ref{source}). The first two terms in the
source term $S$ correspond to an effective potential with a stable
minimum at the self-dual radius. However, if the third term in the
source $S$ does not vanish at the self-dual radius, it will lead to
a positive potential which causes the radion to increase. Thus,
a condition for the stabilization of $b_{\alpha}$ at the self-dual
radius is that the third term in (\ref{source}) vanishes at the
self-dual radius. This is the case if and only if the string state
is a massless mode.

The massless modes have other nice features which are explored in
detail in \cite{Subodh2}. They act as radiation from the
point of view of our three large dimensions and hence do not
lead to a over-abundance problem. As our three spatial dimensions
grow, the potential which confines the radion becomes shallower.
However, rather surprisingly, it turns out the the potential
remains steep enough to avoid fifth force constraints. 

Key to the success in simultaneously avoiding the moduli over-closure
problem and evading fifth force constraints is the fact that
the stabilization mechanism is an intrinsically stringy one.
In the case of a naive effective field theory approach, 
both the confining force and the over-density
in the moduli field scale as $V(\varphi)$, where $V(\varphi)$ is the
potential energy density of the field $\varphi$. In contrast, in
the case of stabilization by means of massless string modes, the
energy density in the string modes (from the point of view of
our three large dimensions) scales as $p_3$, whereas the confining
force scales as $p_3^{-1}$, where $p_3$ is the momentum in the three
large dimensions. Thus, for small values of $p_3$, one simultaneously
gets a large confining force (thus satisfying the fifth force constraints) 
and a small energy density \cite{Subodh2,Subodh3}.

In the presence of massless string states, the shape moduli also
can be stabilized, at least in the simple toroidal backgrounds
considered so far \cite{Edna}. To study this issue, we consider
a metric of the form
\be
ds^2 \, = \, dt^2 - d{\bf x}^2 - G_{mn}dy^mdy^n \, ,
\ee
where the metric of the internal space (here for simplicity
considered to be a two-dimensional torus) contains a shape
modulus, the angle $\theta$ between the two cycles of the torus:
\be
G_{11} \, = \, G_{22} \, = \, 1
\ee
and
\be
G_{12} \, = \, G_{21} \, = \, {\rm sin}\theta \, ,
\ee
where $\theta = 0$ corresponds to a rectangular torus. The ratio
between the two toroidal radii is a second shape modulus. However,
we already know
that each radion individually is stabilized at the self-dual
radius. Thus, the shape modulus corresponding to the ratio of
the toroidal radii is fixed, and the angle is the only shape modulus which has
yet to be considered.

Combining the $00$ and the $12$ Einstein equations, we obtain
a harmonic oscillator equation for $\theta$ with $\theta = 0$
as the stable fixed point.
\be
{\ddot \theta} + 8K^{-1/2} e^{-2 \phi} \theta \, = \, 0 \, ,
\ee
where $K$ is a constant whose value depends on the quantum numbers
of the string gas. In the case of an expanding three-dimensional
space we would have obtained an additional damping term in the
above equation of motion.
We thus conclude that the shape modulus is dynamically stabilized at a value
which maximizes the area to circumference ratio. 

\subsection{Dilaton Stabilization}

The only modulus which is not stabilized with the basic ingredients of
string gas cosmology alone is the dilaton. This situation should be
compared to the problems which arise in the string theory-motivated
approaches to obtaining inflation, where a number of extra ingredients
such as fluxes and non-perturbative effects have to be invoked in order
to stabilize the Kaehler and complex structure moduli (see e.g.
\cite{Eva} for a review).

In string gas cosmology, extra inputs are needed to stabilize the
dilaton. One possibility is that two-loop effective potential effects
can stabilize the dilaton \cite{Cabo}. There have also been
attempts to use extra stringy ingredients such as branes
\cite{Subodh3,Sera,other} or a running tachyon \cite{Kanno1}
to stabilize the dilaton.

The most conservative approach to late-time dilaton stabilization
in string gas cosmology \cite{Frey}, however, is to use one of
the non-perturbative mechanisms which is already widely used
in the literature to fix moduli, namely gaugino condensation
\cite{gaugino}.

Gaugino condensation leads to a correction of the superpotential
$W$ of the theory, from which the actual potential is derived. The
change in the superpotential of the  four-dimensional theory is
\be
W \, \to \, W-Ae^{-1/g^2} \, ,
\ee
where $g$ is the string coupling constant and $A$ is
a constant. The potential $V$
can be derived from the superpotential $W$ and the
Kaehler potential ${\mathcal{K}}$ via the standard formula
\be \label{4dpot}
V \, = \, \frac{1}{M_P^2}e^{\mathcal{K}}\left(\mathcal{K}^{AB}
D_A W D_{{\bar B}}{\bar W} -3|W|^2\right) \ ,
\ee
where the indices $A$ and $B$ run over all of the moduli
fields, and the Kaehler covariant derivative is given by
\be
D_A W \, = \, \partial_A W + (\partial_A \mathcal{K})W \, .
\ee
Since the superpotential in our case is independent of the
volume modulus, the expression for the potential simplifies
to
\be
V \, = \, \frac{1}{M_P^2}e^{\mathcal{K}}\mathcal{K}^{ab}
D_a W D_{\bar b}{\bar W}  \, ,
\ee
where $a$ and $b$ now run only over the modulus
\be
S \, = \, e^{- \Phi} + i a 
\ee
and the complex structure moduli (which we, however,
do not include here). In the above,
$\Phi$ is the four-dimensional dilaton given by
\be
\Phi \, = \, 2 \phi - 6 {\rm ln} b \, ,
\ee
$a$ is the axion, and $M_P$ is the four dimensional Planck
mass.

From the above, we see that the potential 
(\ref{4dpot}) depends both on the dilaton and on
the radion. It is important to verify that adding
this potential to the theory can stabilize the
dilaton without de-stabilizing the radion (which is
fixed by the string gas matter contributions
described in the previous subsection). To investigate
this issue \cite{Frey}, we first need to lift the
potential (\ref{4dpot}) to ten space-time dimensions.
The result, after expanding about the minimum $\Phi_0$
of the potential, is
\bea \label{10dpot}
V(b,\phi) &=& \frac{M_{10}^{16}\hat V}{4} e^{-\Phi_0}
a_0^2 A^2 \left(a_0-\frac{3}{2}e^{\Phi_0}\right)^2
e^{-2a_0 e^{-\Phi_0}}\nonumber\\
&&\times e^{-3\phi/2}\left(b^6 e^{- 2 \phi}-e^{-\Phi_0}\right)^2 \, ,
\eea
where we have written the scale factor in the ten dimensional
Einstein frame. In the above, $\hat V$ is the volume of the
internal space, and $M_{10}$ is the ten dimensional Planck
mass which is given in terms of the volume of the internal
space and the four dimensional Planck mass by
\be
M_P^2 \, = \, M_{10}^8 {\hat V} \, .
\ee
Also, $a_0$ is a constant which appears in the superpotential
(see \cite{Frey} for details).

The effects of gaugino condensation on dilaton and radion
stabilization can now be analyzed in the following way \cite{Frey}:
we start from the dilaton gravity action to which we add the
potential (\ref{10dpot}). To this action we add the action of
a gas of strings, as done in (\ref{dilgrav}). We work in the
ten-dimensional Einstein frame (and thus have to re-scale
the radion, the metric and the matter energy-momentum
tensor accordingly). From this action we can derive the
equations of motion for the dilaton, the radion and the
scale factor of our four-dimensional space-time.

For fixed radion, it follows from (\ref{10dpot}) that
the potential has a minimum for a specific value
of the dilaton. From the considerations of the previous
subsection we know that stringy matter selects
a preferred value of the radion, the self-dual radius.
To demonstrate that the addition of the gaugino potential
can stabilize the dilaton without de-stabilizing the radion
we expand the equations of motion about the value of
the radion corresponding to the self-dual radius and
the value of the dilaton for which the potential
(\ref{10dpot}) is minimized for the chosen value of
the radion. We have shown \cite{Frey} that this is
a stable fixed point of the dynamical system. Thus,
we have shown with the addition of gaugino condensation,
in string gas cosmology all of the moduli are fixed.

\section{String Gas Cosmology and Structure Formation}
 
\subsection{Overview}

At the outset of this section, let us recall the mechanism
by which inflationary cosmology leads to the possibility of
a causal generation mechanism for cosmological fluctuations
which yields an almost scale-invariant spectrum of perturbations.
The space-time diagram of inflationary cosmology is sketched
in Figure 1. In this figure, the vertical axis represents time,
the horizontal axis space (physical as opposed to co-moving
coordinates). The period between times $t_i$ and $t_R$
corresponds to the inflationary phase  (assumed in the figure
to be characterized by almost exponential expansion of space).

During the period of inflation, the Hubble radius
\be
l_H(t) \, = \, \frac{a}{{\dot a}}
\ee
is approximately constant. In contrast, the physical length
of a fixed co-moving scale (labelled by $k$ in the figure)
is expanding exponentially.
In this way, in inflationary cosmology scales which have
microscopic sub-Hubble wavelengths at the beginning of
inflation are red-shifted to become super-Hubble-scale
fluctuations at the end of the period of inflation.
After inflation, the Hubble radius increases linearly in
time, faster than the physical wavelength corresponding
to a fixed co-moving scale. Thus, scales re-enter the
Hubble radius at late times.

The Hubble radius is crucial for the question of generation
of fluctuations for the following reason: If we consider
perturbations with wavelengths smaller than the Hubble radius,
their evolution is dominated by micro-physics which causes
them to oscillate. This is best illustrated by considering the
Klein-Gordon equation for a free scalar field $\varphi$
in an expanding universe. In Fourier space, the equation is
\be \label{KGEOM}
\ddot{\varphi} + 3 H \dot{\varphi} + k_p^2 \varphi \, = \, 0 \, ,
\ee
where $k_p$ is the physical wavenumber. On sub-Hubble
scales $k_p > H$, the Hubble damping term in the above
equation is sub-dominant, and the micro-physics term $k_p^2 \varphi$
leads to oscillations of the field. In contrast, on super-Hubble
scale $k_p < H$, it is the last term on the left-hand side of
(\ref{KGEOM}) which is negligible, and it then follows that
the fluctuations are frozen in.

Thus, if we want to generate primordial cosmological fluctuations
by causal physics, the scale of the fluctuations needs to
be sub-Hubble \footnote{There is, however, a loophole in this
argument: the formation of topological defects during a cosmological
phase transition can lead to non-random entropy fluctuations
on super-Hubble scales which induce cosmological perturbations
in the late universe \cite{VilShell,HK,RHBrev4}.}. In inflationary
cosmology, it is the accelerated expansion of space which enables
the scale of inhomogeneities on current cosmological scales to
be sub-Hubble at early times, and thus leads to the possibility
of a causal generation mechanism for fluctuations.

Since inflation red-shifts any classical fluctuations which might
have been present at the beginning of the inflationary phase,
fluctuations in inflationary cosmology are generated by
quantum vacuum perturbations. The fluctuations begin
in their quantum vacuum state at the onset of inflation. Once the
wavelength exceeds the Hubble radius, squeezing of the
wave-function of the fluctuations sets in (see \cite{MFB,RHBrev2}).
This squeezing plus the de-coherence of the fluctuations due
to the interaction between short and long wavelength modes
generated by the intrinsic non-linearities in both the gravitational and
matter sectors of the theory (see \cite{Martineau,Kiefer,Cliff} for
recent discussions of this aspect and references to previous work)
lead to the classicalization of the fluctuations on super-Hubble
scales.

Let us now turn to the cosmological background of string gas
cosmology represented in Figure 4.  This string gas cosmology
background yields the space-time diagram sketched in Figure 6.
As in Figure 1, the vertical axis is time and 
the horizontal axis denotes the
physical distance. For times $t < t_R$, 
we are in the static Hagedorn phase and the Hubble radius is
infinite. For $t > t_R$, the Einstein frame 
Hubble radius is expanding as in standard cosmology. The time
$t_R$ is when the string winding modes begin to decay into
string loops, and the scale factor starts to increase, leading to the
transition to the radiation phase of standard cosmology. 

\begin{figure}
\begin{center}
%\centerline{\epsfxsize=3in\epsfbox{canc1.eps}}
\includegraphics[height=9cm]{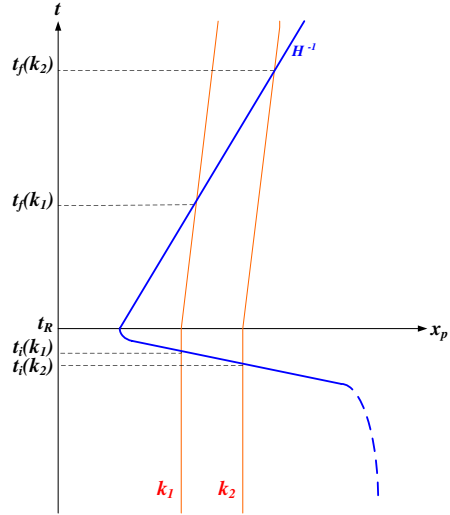}
\caption{Space-time diagram (sketch) showing the evolution of fixed 
co-moving scales in string gas cosmology. The vertical axis is time, 
the horizontal axis is physical distance.  
The solid curve represents the Einstein frame Hubble radius 
$H^{-1}$ which shrinks abruptly to a micro-physical scale at $t_R$ and then 
increases linearly in time for $t > t_R$. Fixed co-moving scales (the 
dotted lines labeled by $k_1$ and $k_2$) which are currently probed 
in cosmological observations have wavelengths which are smaller than 
the Hubble radius before $t_R$. They exit the Hubble 
radius at times $t_i(k)$ just prior to $t_R$, and propagate with a 
wavelength larger than the Hubble radius until they reenter the 
Hubble radius at times $t_f(k)$.}
\end{center}
\label{fig:6}       
\end{figure}

Let us now compare the evolution of the physical 
wavelength corresponding to a fixed co-moving scale  
with that of the Einstein frame Hubble radius $H^{-1}(t)$.
The evolution of scales in string gas cosmology is identical
to the evolution in standard and in inflationary cosmology
for $t > t_R$. If we follow the physical wavelength of the
co-moving scale which corresponds to the current Hubble
radius back to the time $t_R$, then - taking the Hagedorn
temperature to be of the order $10^{16}$ GeV - we obtain
a length of about 1 mm. Compared to the string scale and the
Planck scale, this is a scale in the far infrared.

The physical wavelength is constant in the Hagedorn phase
since space is static. But, as we enter the Hagedorn
phase going back in time, the Hubble radius diverges to
infinity. Hence, as in the case of inflationary cosmology,
fluctuation modes begin sub-Hubble during the Hagedorn
phase, and thus a causal generation mechanism
for fluctuations is possible.

However, the physics of the generation mechanism is
very different. In the case of inflationary cosmology,
fluctuations are assumed to start as quantum vacuum
perturbations because classical inhomogeneities are
red-shifting. In contrast, in the Hagedorn phase of string gas
cosmology there is no red-shifting of classical matter.
Hence, it is the fluctuations in the classical matter which
dominate. Since classical matter is a string gas, the
dominant fluctuations are string thermodynamic fluctuations. 

Our proposal for string gas structure formation is the following 
\cite{NBV} (see \cite{BNPV2} for a more detailed description).
For a fixed co-moving scale with wavenumber $k$ we compute the matter
fluctuations while the scale in sub-Hubble (and therefore gravitational
effects are sub-dominant). When the scale exits the Hubble radius
at time $t_i(k)$ we use the gravitational constraint equations to
determine the induced metric fluctuations, which are then propagated
to late times using the usual equations of gravitational perturbation
theory. Since the scales we are interested
in are in the far infrared, we use the Einstein constraint equations for
fluctuations.

\subsection{String Thermodynamics}

The thermodynamics of a gas of strings was worked out some time
ago \footnote{The initial discussions of the thermodynamics
of strings were given in \cite{Hagedorn,Huang}. More detailed studies
were performed after the first explosion of interest in
superstring theory in the early 1980's \cite{general}.
For some studies of string statistical mechanics particularly relevant
to string gas cosmology see \cite{relevant,Atick}.}. 
We will consider our three spatial dimensions to be compact,
admitting stable winding modes. Specifically, we will take space
to be a three-dimensional torus. In this case, the string gas
specific heat is positive, and string thermodynamics is well-defined,
and was discussed in detail in \cite{Deo} (see also
\cite{Deo2,mt,Turok2,mb}).  What follows is a summary
along the lines of \cite{BNPV2}.

The starting point for our considerations is the free energy $F$
of a string gas in thermal equilibrium
\be \label{free}
F \, = \,- \frac{1}{\beta} ln Z \, ,
\ee
where $\beta$ is the inverse temperature and the canonical partition
function $Z$ is given by
\be \label{part}
Z \, = \, \sum_s e^{-\beta\sqrt{-g_{00}}H(s)} \, ,
\ee
where the sum runs over the states $s$ of the string gas, and $H(s)$ is
the energy of the state. The action $S$ of the string gas is given
in terms of the free energy $F$ via
\be \label{action}
S \, = \, \int dt \sqrt{-g_{00}} F[g_{ij},\beta] \, .
\ee
Note that the free energy depends on the spatial components of the
metric because the energy of the individual string states does.
The energy-momentum tensor $T^{\mu \nu}$ of the string gas is
determined by varying the action with respect to the metric:
\be \label{emtensor}
T^{\mu\nu} \, = \, \frac{2}{\sqrt{-g}}\frac{\delta S}{\delta g_{\mu\nu}} \, .
\ee

Consider now the thermal expectation value
\be \label{texp1}
\langle T^\mu{}_\nu \rangle \, = \,
\frac{1}{Z}\sum_s T^\mu{}_\nu(s)e^{-\beta \sqrt{-g_{00}}H(s)} \, ,
\ee
where $T^\mu{}_\nu (s)$ and $H(s)$ are the energy momentum tensor
and the energy of the state labeled by $s$, respectively. Making use
of (\ref{emtensor}) and (\ref{action}) we immediately find that
\be
T^\mu{}_\nu (s) \, = \, 2 \frac{g^{\mu \lambda}}{\sqrt{-g}}
\frac{\delta }{\delta g^{\lambda \mu}} [-\sqrt{-g_{00}}H(s)] \,,
\ee
and hence
\be \label{texp2}
\langle T^\mu{}_\nu \rangle \, = \, 2 \frac{g^{\mu
\lambda}}{\sqrt{-g}} \frac{\delta \ln{Z}}{\delta g^{\nu\lambda}} \,
.
\ee

To extract the fluctuation tensor of $T_{\mu \nu}$ for long wavelength modes, 
we take one additional variational derivative of (\ref{texp2}), using 
(\ref{texp1}) to obtain
\bea \label{fluct} 
\langle T^\mu{}_\nu T^\sigma{}_\lambda \rangle &
- &\langle T^\mu{}_\nu \rangle \langle T^\sigma{}_\lambda \rangle =
2 \frac{g^{\mu \alpha}}{\sqrt{ - g}}\frac{\partial}{\partial
g^{\alpha\nu}}\left(\frac{g^{\sigma \delta}}{\sqrt{ -
g}}\frac{\partial \ln{Z}}{\partial g^{\delta\lambda}}\right)\nonumber \\
&& + 2 \frac{g^{\sigma \alpha}}{\sqrt{ - g}}\frac{\partial}{\partial
g^{\alpha\lambda}}\left(\frac{G^{\mu \delta}}{\sqrt{ -
g}}\frac{\partial \ln{Z}}{\partial g^{\delta\nu}}\right)\,. 
\eea

As we will see below, the scalar metric fluctuations are
determined by the energy density correlation function
\be
C^0{}_0{}^0{}_0 \, = \, \langle \delta\rho^2 \rangle \, = \,
\langle \rho^2 \rangle - \langle \rho \rangle ^2 \, .
\ee
We will read off the result from the expression (\ref{fluct})
evaluated for $\mu = \nu = \sigma = \lambda = 0$.
The derivative with respect to $g_{00}$ can be expressed in terms of
the derivative with respect to $\beta$. After a couple of steps of
algebra we obtain 
\be \label{cor1}
C^0{}_0{}^0{}_0 \,
= \, - \frac{1}{R^{6}} \frac{\partial}{\partial \beta}
\left(F + \beta \frac{\partial F}{\partial \beta}\right) \, = \,
\frac{T^2}{R^6} C_V \, .
\ee
where
\be \label{specheat} C_V \, = \, (\partial  E  /
\partial T)|_{V} \,,
\ee
is the specific heat, and
\be
E \, \equiv \, F + \beta \left(\frac{\partial F}{\partial
\beta}\right) \,,
\ee
is the internal energy. Also, $V = R^3$ is the
volume of the three compact but large spatial dimensions.

The gravitational waves are determined by the off-diagonal
spatial components of the correlation function tensor, i.e.
\be \label{cor2a} 
C^i{}_j{}^i{}_j \, = \, \langle \delta {T^i{}_j}^2
\rangle \, = \, \langle {T^i{}_j}^2 \rangle - \langle T^i{}_j
\rangle^2\,,
\ee
with $i \neq j$. 

Our aim is to calculate the fluctuations of the energy-momentum
tensor on various length scales $R$. For each value of $R$, we will
consider string thermodynamics in a box in which all edge lengths
are $R$. From (\ref{fluct}) it is obvious that in order to have
non-vanishing off-diagonal spatial correlation functions, we must
consisted a torus with its shape moduli turned on. Let us focus on the 
$x-y$ component of
the correlation function. We will consider the spatial part of the
metric to be 
\be \label{metric2}
ds^2 \, = \, R^2 d\theta_x^2 + 2 \epsilon R^2 d\theta_x d\theta_y
+ R^2 d\theta_y^2
\ee
with $\epsilon \ll 1$. The spatial coordinates $\theta_i$ run over a
fixed interval, e.g. $[0, 2\pi]$), The generalization of the spatial
part of the metric to three dimensions is obvious. At the end of the
computations, we will set $\epsilon = 0$.

{F}rom the form of (\ref{fluct}), it follows that all
space-space correlation function tensor elements are of the same
order of the magnitude, namely 
\be \label{cor2}
C^i{}_j{}^i{}_j
\, = \, \frac{1}{\beta R^3}\frac{\partial}{\partial \ln{R}}\left(-
\frac{1}{R^3} \frac{\partial F}{\partial \ln{R}}\right)  =
\frac{1}{\beta R^2}\frac{\partial p}{\partial R} \, ,
\ee
where the string pressure is given by
\be
p \,  \equiv   \, -\frac{1}{V}\left(\frac{\partial F}{\partial
\ln{R}}\right) \, = \, T \left(\frac{\partial S}{\partial
V}\right)_E \,. \label{stringypressure}
\ee

In the following, we will compute the two correlation functions
(\ref{cor1}) and (\ref{cor2}) using tools from string statistical
mechanics. Specifically, we will be following the discussion in
\cite{Deo}. The starting point is the formula
\be
S(E , R) \, = \, \ln{\Omega(E ,R)}
\ee
for the entropy in terms of $\Omega(E ,R)$, the density of states.
The density of states of a gas of closed strings on a large
three-dimensional torus (with the radii of all internal dimensions
at the string scale) was calculated in \cite{Deo} (see also
\cite{Ali3}) and is given by
\be
\Omega(E , R) \, \simeq \, \beta_H e^{\beta_H E + n_H V}[1 +
\delta \Omega_{(1)}(E , R)] \label{density_states}\,,
\ee
where $\delta \Omega_{(1)}$ comes from the contribution to the
density of states (when writing the density of states as an 
inverse Laplace
transform of $Z(\beta)$, which involves integration over $\beta$)
from the closest singularity point $\beta_1$ to $\beta_H = (1/T_H)$
in the complex $\beta$ plane. Note that $\beta_1 < \beta_H$, and
$\beta_1$ is real. From \cite{Deo,Ali3} we have
\be \label{deltaomega0}
\delta \Omega_{(1)}(E , R) \, = \, - \frac{(\beta_H E)^{5}}{5!}
e^{-(\beta_H - \beta_1)(E  - \rho_H V)} \,.
\ee
In the above, $n_H$ is a (constant) number density of order
$l_s^{-3}$  and $\rho_H$ is the `Hagedorn Energy density' of the
order $l_s^{-4}$, and
\be
\beta_H - \beta_1 \sim \left\{ \begin{array}{ll}
(l_s^3/R^2) \,, & \mbox{for $R \gg l_s$}\,, \\
(R^2/l_s)\,, & \mbox{for $R \ll l_s$}\,.
\end{array}
\right.
\ee
To ensure the validity of Eq. (\ref{density_states}) we demand that
\be \label{cond2}
- \delta \Omega_{(1)} \,\ll \, 1
\ee
by assuming $\rho \equiv (E / V) \gg \rho_H$, which corresponds to being
in a state in which winding modes and oscillatory modes can be excited
and we expect important deviations from point particle thermodynamics.

Combining the above results, we find that the entropy of the string gas
in the Hagedorn phase is given by
\be
\label{entropy} S(E , R) \simeq \beta_H E + n_H V + \ln{\left[1
+ \delta \Omega_{(1)}\right]} \,,
\ee
and therefore the temperature $T (E, R) \equiv [(\partial S/\partial
E)_V]^{-1}$ will be
\bea
T  &\simeq&  \left(\beta_H + \frac{\partial \delta
\Omega_{(1)}/\partial E}{1 + \delta \Omega_{(1)}}\right)^{-1} \nonumber \\
&\simeq&
T_H \left(1 + \frac{\beta_H - \beta_1}{\beta_H} \delta
\Omega_{(1)}\right)\label{temp}\,.
\eea
In the above, we have dropped a term which is negligible since
$E (\beta_H - \beta_1) \gg 1$ (see (\ref{cond2})).
Using this relation, we can express $\delta \Omega_{(1)}$ in terms of
$T$ and $R$ via
\be \label{deltaomega}
l_s^3\delta \Omega_{(1)} \, \simeq \, -
\frac{R^2}{T_H}\left(1 - \frac{T}{T_H}\right) \, .
\ee
In addition, we find
\be \label{meanen}
E  \, \simeq \,  l_s^{-3} R^2
\ln{\left[\frac{\ell_s^3 T}{R^2 (1- T/T_H)}\right]}\,.
\ee
Note that to ensure that $|\delta \Omega_{(1)}| \ll 1$ and $E \gg
\rho_H R^3$, one should demand %%
\be \label{cond}
(1 - T/T_H) R^2 l_s^{-2} \ll 1 \, .
\ee

The results (\ref{entropy}) and (\ref{deltaomega}) now allow us
to compute the correlation functions (\ref{cor1}) and (\ref{cor2}).
We first compute the energy correlation function (\ref{cor1}).
Making use of (\ref{meanen}), it follows from
(\ref{specheat}) that
\be \label{specheat2}
C_V  \, \approx \,  \frac{R^2/l_s^3}{T \left(1
- T/T_H\right)} \,,
\ee
from which we get
\be 
C^0{}_0{}^0{}_0 = \langle\delta \rho^2\rangle\, \simeq \,
\frac{T}{l_s^3(1 - T/T_H)} \frac{1}{R^4}\, . 
\ee
Note that the factor $(1 - T/T_H)$ in the denominator will
turn out to be responsible for giving the 
spectrum a slight red tilt. It comes from the differentiation
with respect to $T$.

Next we evaluate (\ref{stringypressure}). From the definition of the
pressure it follows that (to linear order in $\delta \Omega_{(1)}$)
\be
p \, = \, n_H T + T {{\partial} \over {\partial V}} \delta \Omega_{(1)}
\, ,
\ee
where the final partial derivative is to be taken at constant
energy. In taking this partial derivative, we insert the
expression (\ref{deltaomega}) for $\delta \Omega_{(1)}$ and must keep
careful account of the fact that $\beta_H - \beta_1$ depends
on the radius $R$. In evaluating the resulting terms, we
keep only the one which dominates at high energy density. It
is the term which comes from differentiating the factor
$\beta_H - \beta_1$. This differentiation brings down a
factor of $E$, which is then substituted by means
of (\ref{meanen}), thus introducing a logarithmic factor
in the final result for the pressure. We obtain
\be
p(E, R) \, \approx \, n_H T_H - \frac{2}{3}\frac{(1 - T/T_H)}{l_s^3
R}\ln{\left[\frac{l_s^3 T}{R^2 (1- T/T_H)}\right]} \,,
\ee
which immediately yields
\be
\label{tensorresult} C^i{}_j{}^i{}_j \, \simeq \, \frac{T (1 -
T/T_H)}{l_s^3 R^4} \ln^2{\left[\frac{R^2}{l_s^2}(1 -
T/T_H)\right]}\, .
\ee
Note that since no temperature derivative is taken, the factor 
$(1 - T/T_H)$ remains in the numerator. This is one of the 
two facts which will lead to the
slight blue tilt of the spectrum of gravitational waves. The second
factor contributing to the slight blue tilt is the explicit factor
of $R^2$ in the logarithm. Because of (\ref{cond}), we are on the
large $k$ side of the zero of the logarithm. Hence, the greater the
value of $k$, the larger the absolute value of the logarithmic factor.

\subsection{Spectrum of Cosmological Fluctuations}

We write the metric including cosmological perturbations
(scalar metric fluctuations) and gravitational waves in the
following form:
\be \label{pertmetric}
d s^2 \, = \, a^2(\eta) \left\{(1 + 2 \Phi)d\eta^2 - [(1 - 
2 \Phi)\delta_{ij} + h_{ij}]d x^i d x^j\right\} \, . 
\ee 
In the above, we have used conformal time $\eta$ which is related to
the physical time $t$ via
\be
dt \, = \, a(t) d\eta \, .
\ee
We have fixed the gauge (i.e. coordinate) freedom for the scalar metric
perturbations by adopting the longitudinal gauge in terms of which
the metric is diagonal. Furthermore, we have taken matter to be free
of anisotropic stress (otherwise there would be two scalar metric degrees
of freedom instead of the single function $\Phi({\bf x}, t)$). The spatial
tensor $h_{ij}({\bf x}, t)$ is transverse and traceless and represents the
gravitational waves. 

Note that in contrast to the case of slow-roll inflation, scalar metric
fluctuations and gravitational waves are generated by matter
at the same order in cosmological perturbation theory. This could
lead to the expectation that the amplitude of gravitational waves
in string gas cosmology could be generically larger than in inflationary
cosmology. This expectation, however, is not realized \cite{BNPV1}
since there is a different mechanism which suppresses the gravitational
waves relative to the density perturbations (namely the fact
that the gravitational wave amplitude is set by the amplitude of
the pressure, and the pressure is suppressed relative to the
energy density in the Hagedorn phase).

Assuming that the fluctuations are described by the perturbed Einstein
equations (they are {\it not} if the dilaton is not fixed 
\cite{Betal,KKLM}), then the spectra of cosmological perturbations
$\Phi$ and gravitational waves $h$ are given by the energy-momentum 
fluctuations in the following way \cite{BNPV2}
\be \label{scalarexp} 
\langle|\Phi(k)|^2\rangle \, = \, 16 \pi^2 G^2 
k^{-4} \langle\delta T^0{}_0(k) \delta T^0{}_0(k)\rangle \, , 
\ee 
where the pointed brackets indicate expectation values, and 
\be 
\label{tensorexp} \langle|h(k)|^2\rangle \, = \, 16 \pi^2 G^2 
k^{-4} \langle\delta T^i{}_j(k) \delta T^i{}_j(k)\rangle \,, 
\ee 
where on the right hand side of (\ref{tensorexp}) we mean the 
average over the correlation functions with $i \neq j$, and
$h$ is the amplitude of the gravitational waves \footnote{The
gravitational wave tensor $h_{i j}$ can be written as the
amplitude $h$ multiplied by a constant polarization tensor.}.
 
Let us now use (\ref{scalarexp}) to determine the spectrum of
scalar metric fluctuations. We first calculate the 
root mean square energy density fluctuations in a sphere of
radius $R = k^{-1}$. For a system in thermal equilibrium they 
are given by the specific heat capacity $C_V$ via 
(see (\ref{cor1})
\be \label{cor1b}
\langle \delta\rho^2 \rangle \,  = \,  \frac{T^2}{R^6} C_V \, . 
\ee 
From the previous subsection we know that the specific 
heat of a gas of closed strings
on a torus of radius $R$ is (see \ref{specheat2})
\be \label{specheat2b} 
C_V  \, \approx \, 2 \frac{R^2/\ell^3}{T \left(1 - T/T_H\right)}\, . 
\ee 
Hence, the power spectrum $P(k)$ of scalar metric fluctuations can
be evaluated as follows
\bea \label{power2} 
P_{\Phi}(k) \, & \equiv & \, {1 \over {2 \pi^2}} k^3 |\Phi(k)|^2 \\
&=& \, 8 G^2 k^{-1} <|\delta \rho(k)|^2> \, . \nonumber \\
&=& \, 8 G^2 k^2 <(\delta M)^2>_R \nonumber \\ 
               &=& \, 8 G^2 k^{-4} <(\delta \rho)^2>_R \nonumber \\
&=& \, 8 G^2 {T \over {\ell_s^3}} {1 \over {1 - T/T_H}} 
\, , \nonumber 
\eea 
where in the first step we have used (\ref{scalarexp}) to replace the 
expectation value of $|\Phi(k)|^2$ in terms of the correlation function 
of the energy density, and in the second step we have made the 
transition to position space 

The first conclusion from the result (\ref{power2}) is that the spectrum
is approximately scale-invariant ($P(k)$ is independent of $k$). It is
the `holographic' scaling $C_V(R) \sim R^2$ which is responsible for the
overall scale-invariance of the spectrum of cosmological perturbations.
However, there is a small $k$ dependence which comes from the fact
that in the above equation for a scale $k$ 
the temperature $T$ is to be evaluated at the
time $t_i(k)$. Thus, the factor $(1 - T/T_H)$ in the 
denominator is responsible 
for giving the spectrum a slight dependence on $k$. Since
the temperature slightly decreases as time increases at the
end of the Hagedorn phase, shorter wavelengths for which
$t_i(k)$ occurs later obtain a smaller amplitude. Thus, the
spectrum has a slight red tilt.

\subsection{Spectrum of Gravitational Waves}

As discovered in \cite{BNPV1}, the spectrum of gravitational
waves is also nearly scale invariant. However, in the expression
for the spectrum of gravitational waves the factor $(1 - T/T_H)$
appears in the numerator, thus leading to a slight blue tilt in
the spectrum. This is a prediction with which the cosmological
effects of string gas cosmology can be distinguished from those
of inflationary cosmology, where quite generically a slight red
tilt for gravitational waves results. The physical reason is that
large scales exit the Hubble radius earlier when the pressure
and hence also the off-diagonal spatial components of $T_{\mu \nu}$
are closer to zero.

Let us analyze this issue in a bit more detail and
estimate the dimensionless power spectrum of gravitational waves.
First, we make some general comments. In slow-roll inflation, to
leading order in perturbation theory matter fluctuations do not
couple to tensor modes. This is due to the fact that the matter
background field is slowly evolving in time and the leading order
gravitational fluctuations are linear in the matter fluctuations. In
our case, the background is not evolving (at least at the level of
our computations), and hence the dominant metric fluctuations are
quadratic in the matter field fluctuations. At this level, matter
fluctuations induce both scalar and tensor metric fluctuations.
Based on this consideration we might expect that in our string gas
cosmology scenario, the ratio of tensor to scalar metric
fluctuations will be larger than in simple slow-roll inflationary
models. However, since the amplitude $h$ of the gravitational
waves is proportional to the pressure, and the pressure is suppressed
in the Hagedorn phase, the amplitude of the gravitational waves
will also be small in string gas cosmology.

The method for calculating the spectrum of gravitational waves
is similar to the procedure outlined in the last section
for scalar metric fluctuations. For a mode with fixed co-moving
wavenumber $k$, we compute the correlation function of the
off-diagonal spatial elements of the string gas energy-momentum
tensor at the time $t_i(k)$ when the mode exits the Hubble radius
and use (\ref{tensorexp}) to infer the amplitude of the power
spectrum of gravitational waves at that time. We then
follow the evolution of the gravitational wave power spectrum
on super-Hubble scales for $t > t_i(k)$ using the equations
of general relativistic perturbation theory.

The power spectrum of the tensor modes is given by (\ref{tensorexp}):
\be \label{tpower1}
P_h(k) \, = \, 16 \pi^2 G^2 k^{-4} k^3
\langle\delta T^i{}_j(k) \delta T^i{}_j(k)\rangle
\ee
for $i \neq j$. Note that the $k^3$ factor multiplying the momentum
space correlation function of $T^i{}_j$ gives the position space
correlation function, namely the root mean square fluctuation of
$T^i{}_j$ in a region of radius $R = k^{-1}$ (the reader who is
skeptical about this point is invited to check that the dimensions
work out properly). Thus,
\be \label{tpower2}
P_h(k) \, = \, 16 \pi^2 G^2 k^{-4} C^i{}_j{}^i{}_j(R) \, .
\ee
The correlation function $C^i{}_j{}^i{}_j$ on the right hand side
of the above equation was computed earlier (in the subsection
on string thermodynamics). Inserting
the result (\ref{tensorresult}) into (\ref{tpower1}) we obtain
\be \label{tpower3}
P_h(k) \, \sim \, 16 \pi^2 G^2 {T \over {l_s^3}}
(1 - T/T_H) \ln^2{\left[\frac{1}{l_s^2 k^2}(1 -
T/T_H)\right]}\, ,
\ee
which, for temperatures close to the Hagedorn value reduces to
\be \label{tresult}
P_h(k) \, \sim \,
\left(\frac{l_{Pl}}{l_s}\right)^4 (1 -
T/T_H)\ln^2{\left[\frac{1}{l_s^2 k^2}(1 - T/T_H)\right]} \, .
\ee
This shows that the spectrum of tensor modes is - to a first
approximation, namely neglecting the logarithmic factor and
neglecting the $k$-dependence of $T(t_i(k))$ - scale-invariant. The
corrections to scale-invariance will be discussed at the end of this
subsection.

On super-Hubble scales, the amplitude $h$ of the gravitational waves
is constant. The wave oscillations freeze out when the wavelength
of the mode crosses the Hubble radius. As in the case of scalar metric
fluctuations, the waves are squeezed. Whereas the wave amplitude remains
constant, its time derivative decreases. Another way to see this
squeezing is to change variables to 
\be
\psi(\eta, {\bf x}) \, = \, a(\eta) h(\eta, {\bf x})
\ee
in terms of which the action has canonical kinetic term. The action
in terms of $\psi$ becomes
\be
S \, = \, {1 \over 2} \int d^4x \left( {\psi^{\prime}}^2 -
\psi_{,i}\psi_{,i} + {{a^{\prime \prime}} \over a} \psi^2 \right)
\ee
from which it immediately follows that on super-Hubble scales
$\psi \sim a$. This is the squeezing of gravitational 
waves \cite{Grishchuk}.

Since there is no $k$-dependence in the squeezing factor, the
scale-invariance of the spectrum at the end of the Hagedorn phase
will lead to a scale-invariance of the spectrum at late times.

Note that in the case of string gas cosmology, the squeezing
factor $z(\eta)$ does not differ substantially from the
squeezing factor $a(\eta)$ for gravitational waves. In the
case of inflationary cosmology, $z(\eta)$ and $a(\eta)$
differ greatly during reheating, leading to a much larger
squeezing for scalar metric fluctuations, and hence to a
suppressed tensor to scalar ratio of fluctuations. In the
case of string gas cosmology, there is no difference in
squeezing between the scalar and the tensor modes. 

Let us return to the discussion of the spectrum of gravitational
waves. The result for the power spectrum is given in
(\ref{tresult}), and we mentioned that to a first approximation this
corresponds to a scale-invariant spectrum. As realized in
\cite{BNPV1}, the logarithmic term and the $k$-dependence of
$T(t_i(k))$ both lead to a small blue-tilt of the spectrum. This
feature is characteristic of our scenario and cannot be reproduced
in inflationary models. In inflationary models, the amplitude of
the gravitational waves is set by the Hubble constant $H$. The
Hubble constant cannot increase during inflation, and hence no
blue tilt of the gravitational wave spectrum is possible.

To study the tilt of the tensor spectrum, we first have to keep in
mind that our calculations are only valid in the range (\ref{cond}),
i.e. to the large $k$ side of the zero of the logarithm. Thus, in
the range of validity of our analysis, the logarithmic factor
contributes an explicit blue tilt of the spectrum. The second source
of a blue tilt is the factor $1 - T(t_i(k)) / T_H$ multiplying the
logarithmic term in (\ref{tresult}). Since modes with larger values
of $k$ exit the Hubble radius at slightly later times $t_i(k)$, when
the temperature $T(t_i(k))$ is slightly lower, the factor will be
larger. 

A heuristic way of understanding the origin of the slight blue tilt
in the spectrum of tensor modes
is as follows. The closer we get to the Hagedorn temperature, the
more the thermal bath is dominated by long string states, and thus
the smaller the pressure will be compared to the pressure of a pure
radiation bath. Since the pressure terms (strictly speaking the
anisotropic pressure terms) in the energy-momentum tensor are
responsible for the tensor modes, we conclude that the smaller the
value of the wavenumber $k$ (and thus the higher the temperature
$T(t_i(k))$ when the mode exits the Hubble radius, the lower the
amplitude of the tensor modes. In contrast, the scalar modes are
determined by the energy density, which increases at $T(t_i(k))$ as
$k$ decreases, leading to a slight red tilt.

\subsection{Discussion}

To summarize this section, we have seen that string gas cosmology
provides a mechanism alternative to the well-known inflationary
one for generating an approximately scale-invariant spectrum of
approximately adiabatic density fluctuations. The model predicts
a slight red tilt of the spectrum (as is also obtained in many
simple inflationary models). However, as a prediction which 
distinguishes the model from the inflationary universe scenario,
it predicts a slight blue tilt of the spectrum of gravitational
waves. Thus, a way to rule out the inflationary scenario would
be to detect a stochastic background of gravitational waves at 
both very small wavelengths (using direct detection experiments
such as gravitational wave antennas) and on cosmological scales
(using signatures in CMB temperature maps) and to infer
a blue tilt from those measurements. The current limits
on the magnitude of the blue tilt are not very strong \cite{Stewart}
but can be improved.

The scenario has one free parameter (the ratio of the string
to the Planck length) and one free function (the k-dependence
of the temperature $T(t_i(k))$ - in principle calculable if
the dynamics of the exit from the Hagedorn phase were better
known). There are five basic observables: the amplitudes of
the scalar and tensor spectra, their tilts, and the amplitude
of the jump in the CMB temperature maps produced by long
straight cosmic superstrings via the Kaiser-Stebbins \cite{KS}
effect. Thus, there are three consistency relations between the
observables which allow the scenario to be falsified.

An important point is that the thermal string gas fluctuations evolve
for a long time during the radiation phase outside the Hubble radius.
Like in inflationary cosmology, this
leads to the squeezing of fluctuations which is responsible for
the acoustic oscillations in the angular power spectrum of CMB
anisotropies (see e.g. \cite{BNPV2} for a more detailed discussion
of this point). Note that the situation is completely different
from that in topological defect models of structure formation,
where the curvature perturbations are constantly seeded from the
defect sources at late times, and which hence does not lead
to the oscillations in the angular power spectrum of the CMB.

The non-Gaussianities induced by the thermal gas of strings
are large on microscopic scale, but Poisson-suppressed on
larger scales. The three point correlation function produced
by a string gas and the related non-Gaussianity parameter
can be calculated \cite{Chen} from the same starting point
of string thermodynamics described earlier in this section.

Note that the structure formation scenario discussed in this
section relies on three key assumptions - firstly the holographic
scaling of the specific heat capacity $C_V(R) \sim R^2$, secondly
the applicability of the Einstein equations to describe metric
fluctuations on infrared scales, and thirdly the
existence of a phase like the Hagedorn phase at the end of
which the matter fluctuations seed metric perturbations (it may
be a phase only describable using a truly non-perturbative
string theory or quantum gravity framework). For attempts
to realize our basic structure formation scenario in a
different context see e.g \cite{Joao}. 

Dilaton gravity does not provide a satisfactory framework to
implement our structure formation scenario \cite{Betal,KKLM} and
is unsatisfactory as a description of the Hagedorn phase for
other reasons mentioned earlier in this chapter. The
criticisms of string gas cosmology in \cite{KKLM,Kaloper}
are mostly problems which are specific to the attempt to use
dilaton gravity as the background for string gas cosmology.
There is a specific background model in which all of the key assumptions
discussed above are realized, namely the ghost-free higher derivative
gravity action of \cite{BMS} which yields a non-singular bouncing
cosmology. If we add string gas matter to this action and adjust
parameters of the model such that the bounce phase is long, then
thermal string fluctuations in the bounce phase yield a
realization of our scenario \cite{BBMS}. 

\section{Conclusions}

The string gas scenario is an approach to early universe cosmology based
on coupling a gas of strings to a classical background. It includes
string degrees of freedom and string symmetries which are hard to implement
in an effective field theory approach.

The background of string gas cosmology is non-singular. The temperature
never exceeds the limiting Hagedorn temperature. If we start the evolution
as a dense gas of strings in a space in which all dimensions are string-scale
tori, then there are dynamical arguments according to which only three
of the spatial dimensions can become large \cite{BV}. Thus, string gas
cosmology yields the hope of understanding why - in the context of
a theory with more than three spatial dimensions - exactly three are
large and visible to us.

If the Hagedorn phase (the phase during which the temperature is close
to the Hagedorn temperature and both the scale factor and the dilaton
are static) is sufficiently long to establish thermal equilibrium
on length scales of about 1 mm, then string gas cosmology can provide
an alternative to cosmological inflation for explaining the origin of
an almost scale-invariant spectrum of cosmological fluctuations \cite{NBV}. A
distinctive signature of the scenario is the slight blue tilt in the
spectrum of gravitational waves which is predicted \cite{BNPV1}.

The inflationary universe scenario has successes beyond the 
fact that it successfully predicted a scale-invariant spectrum
of fluctuations - it also explains why, starting from a hot
Planck scale space, an extremely low entropy state - one
can obtain a universe which is large enough and contains
enough entropy to correspond to our observed universe. In addition,
it explains the observed spatial flatness. However, if the Hagedorn
phase of string gas cosmology is realized as a long bounce phase
in a universe which starts out large and cold, then the horizon,
flatness, size and entropy problems do not arise.

A serious concern for the current realization of string gas 
cosmology, however, is the gravitational 
Jeans instability problem. This problem was first
raised in \cite{stability}. In the context of dilaton gravity,
it can be shown \cite{dilflucts} that gravitational fluctuations
do not grow. However, dilaton gravity is not a consistent background
for the Hagedorn phase of string gas cosmology. One might hope
that since the string states are relativistic, the gravitational
Jeans length will be comparable to the Hubble radius, as it is
for a gas of regular radiation. However, a recent computation of
the speed of sound in string gas cosmology \cite{Nima} has shown
that in a background space sufficiently large to evolve into
our present universe the overall speed of sound is very small.
Further work needs to be done on this issue. This is
complicated by fact that string thermodynamics 
is non-extensive (see e.g. \cite{Cobas2}), which
leads to problems in using the usual thermodynamic intuition.

Note that the background space does not need to be toroidal. Crucial for 
string gas cosmology
to yield the predictions summarized above is the existence and stability
(or quasi-stability) of string winding modes. Certain orbifolds 
\cite{Col1} have also been shown to yield good backgrounds for string
gas cosmology. Non-trivial one cycles will ensure the existence
and stability of string winding modes. 

If the background space does not have have any non-trivial one-cycles,
then it might be possible to construct a cosmological scenario based
on stable branes rather than strings. The cosmology of brane gases
has been considered in \cite{branes}. If there are stable winding
strings, and if the string coupling constant is small such that
the fundamental strings are lighter than branes, then \cite{ABE}
it is the fundamental strings which will dominate the thermodynamics
in the Hagedorn phase and which will be the most important degrees
of freedom for cosmology. However, if there are no stable winding
strings, then winding branes would become important.

It appears at the present time that Heterotic string theory is most
suited for string gas cosmology since this theory admits the enhanced
symmetry states which have been shown to yield a very simple way
to stabilize the size and shape moduli of the extra spatial dimensions.
It will be interesting to study if string gas cosmology can be
embedded into particular models of Heterotic string theory which
yield reasonable particle phenomenology.

The presentation we gave of string gas cosmology is based on minimal
input. In particular, we did not include fluxes since we assume that
the net fluxes should cancel for a situation with the most symmetric
initial conditions. The role of fluxes in string gas cosmology has been
studied in \cite{Campos}. Whereas the primary application of string
gas cosmology will be to the cosmology of the very early universe,
it is also interesting to consider applications of string gas cosmology
to later time cosmology. The late time dynamics of string and brane
gases has been considered in \cite{late}. In particular, in \cite{Ferrer}
applications of string gases to the dark energy problem has
been considered (see also \cite{McInnes}). String and brane gases have also
been studied as a way to obtain inflation \cite{Turok,Parry,Anupam} (see
also \cite{Freese}), or
as a way to obtain non-inflationary bulk expansion which may provide
a way to solve the size problem in string gas cosmology if one starts with
a spatial manifold of string scale in all directions \cite{Natalia}. 

\vskip0.3cm

\centerline{\bf Acknowledgements}

\vskip0.2cm

I am grateful to all of my present and former collaborators
with whom I have had the pleasure of working on string gas
cosmology. For comments on the draft of this manuscript I
wish to thank Nima Lashkari and Subodh Patil. This work is
supported in part by an NSERC Discovery Grant and by funds
from the Canada Research Chairs Program.

\end{document}